\documentclass[onecolumn]{aa}
\usepackage{graphicx}
\usepackage{txfonts}
\begin{document}
   \title{The mass loss of C-rich giants \thanks{This research has made use of the Simbad database 
operated at CDS}\fnmsep\thanks{Partially based on data from the ESA HIPPARCOS astrometry satellite}
\fnmsep\thanks{Table 3 is only available in electronic form at the CDS via anonymous ftp 130.79.128.5}}


   \author{J. Bergeat
          \and
          L. Chevallier 
          }

   \offprints{J. Bergeat}

   \institute{Centre de Recherche Astronomique de Lyon (UMR 5574 du CNRS), Observatoire de
             Lyon,\\ 
9 avenue Charles Andr\'e, 69561 St-Genis-Laval cedex, France
             }

   \date{Received 12 May 2004 / Accepted 23 July 2004}

   \abstract{
   The mass loss rates, expansion velocities and dust-to-gas density ratios from millimetric observations
   of 119 carbon-rich giants are compared, as functions of stellar parameters, to the predictions of recent 
   hydrodynamical models. Distances and luminosities previously estimated from HIPPARCOS data, masses from
   pulsations and C/O abundance ratios from spectroscopy, and effective temperatures from a new homogeneous
   scale, are used. Predicted and observed mass loss rates agree fairly well, as functions of effective 
   temperature. The signature of the mass range $\rm{M\le 4\,M_{\sun}}$ of most carbon-rich AGB stars is 
   seen as a flat portion in the diagram of mass loss rate {\it vs.} effective temperature.
   It is flanked by two regions of mass loss rates increasing with decreasing effective temperature at
   nearly constant stellar mass. Four stars with detached shells, i.e. episodic strong mass loss, and
   five cool infrared carbon-rich stars with optically-thick dust shells, have mass loss rates much larger
   than predicted values. The latter (including CW Leo) could be stars of smaller masses 
   ($\rm{M\simeq 1.5-2.5\,M_{\sun}}$) while $\rm{M\simeq 4\,M_{\sun}}$ is indicated for most of the 
   coolest objects. Among the carbon stars with detached shells, R Scl returned to a predicted
   level (16 times lower) according to recent measurements of the central source. 
   The observed expansion velocities are in agreement with the predicted velocities at infinity in a 
   diagram of velocities {\it vs.} effective temperature, provided the carbon to oxygen 
   abundance ratio is $\rm{1\le \epsilon _{C}/\epsilon _{O}\le 2},$ i.e. the range 
   deduced from spectra and model atmospheres of those cool variables. Five stars with detached shells
   display expansion velocities about twice that predicted at their effective temperature. Miras and
   non-Miras do populate the same locus in both diagrams at the present accuracy. The predicted dust-to-gas
   density ratios are however about 2.2 times smaller than the values estimated from observations. Recent 
   drift models can contribute to minimize the discrepancy since they include more dust. Simple 
   approximate formulae are proposed. 
   \keywords{stars: AGB and post-AGB -- stars: carbon -- stars: mass loss
               }
   }
\twocolumn
   \maketitle
%

\section{Introduction}

   The observations of CO and HCN lines in the millimetric range together with infrared fluxes, especially 
   from the IRAS catalogue (\cite{IRAS88}), provide estimates of parameters of the circumstellar 
   envelopes of carbon-rich giants. They are essentially TP-AGB stars, i.e. late stages of stellar
   evolution of low and intermediate mass stars, experiencing substantial mass loss.
   This latter phenomenon may terminate their evolution on the AGB before rapid nucleosynthesis forces
   them to leave that branch. Mass loss also contributes to the replenishment of the interstellar medium 
   in carbonaceous material (dust and gas), mostly from extreme (``infrared'') carbon stars. Models and
   simplifying assumptions are required to derive the estimates of the mass loss rate 
   $\rm{\dot{M}}$ in solar mass per year,
   the expansion velocity $\rm{v_{e}}$ in kilometer per second, and the dust-to-gas mass density ratio.
   For instance a model for the photodissociation of the circumstellar CO 
   by Mamon et al. (\cite{mamon88}) is usually applied to radial brightness distributions with fair
   success, but clear deviations were noted in a few cases (e.g. Sch\"{o}ier \& Olofsson \cite{schoie01}). 
   Predominant species, like hydrogen or helium, are not traced out and the adopted abundances influence 
   the estimates of the total mass loss rate. The estimated distances are also of paramount importance
   since the rate calculated from observations increases as the squared distance. A full discussion
   is outside the scope of the present paper and we refer the reader to the papers cited in Sect. 2.

   It has become clear that the interaction of pulsation and dust formation plays a key role in the mass
   loss phenomenon (e.g. Wallerstein \& Knapp \cite{waller98} for a review). The effects of stellar
   pulsation are simulated in dynamical models by applying a piston at the inner boundary. Consistent 
   models of circumstellar 
   dust shells around carbon-rich long period variables (LPV) include time-dependent hydrodynamics, a
   detailed treatment of the processes of formation, growth and evaporation of dust grains, a carbon-rich
   chemistry, and radiative transfer (e. g. Fleischer et al. \cite{fleisc92}, Fleischer \cite{fleisc94},
   Winters et al. \cite{winter94}, \cite{winter95}, H\"{o}fner et al. \cite{hofner95}, \cite{hofner96}).
   The empirical Reimers relation for RGB (Reimers \cite{reimer75}) proved its inadequacy for AGB. 
   Alternative
   relations were derived from observational results and/or theoretical models of mass loss (e.g. Volk
   \& Kwok \cite{volk88}, Vassiliadis \& Wood \cite{vassi93}, Bl\"{o}cker \cite{blocke95}). The mass loss 
   relation of Dominik et al. (\cite{domini90}) was the first one based on detailed calculations of dust 
   formation and growth (Gail \& Sedlmayr \cite{gail88}), with rather extreme stellar parameters adapted
   to the very late stages of AGB evolution. 
   Approximate formulae for mass loss rate, expansion velocity and dust to gas density ratio were then
   calculated by Arndt et al. (\cite{arndt97}). They considered six stellar parameters as independent
   ones, namely the temperature, the luminosity, the mass, the abundance ratio of carbon to oxygen, the
   pulsational period and the velocity amplitude of the pulsation. They then derived approximate equations
   on the basis of 48 dynamical models, exploring various ranges for the six parameters. They also
   compared their results to those of 22 analogous models from H\"{o}fner \& Dorfi (\cite{hofner97})
   who proved to be well approximated by their set of equations.
 \begin{table*}
 \caption
 {Mean distances (pc) of carbon-rich giants and dispersions as obtained from data of various authors:
 O93 stands for Olofsson et al. (\cite{olofs93a}), G99 for Groenewegen et al. (\cite{groene99}),
 SO1 for Sch\"{o}ier \& Olofsson (\cite{schoie01}), GO2 for Groenewegen et al. (\cite{groen02b}),
 and L93 for Loup et al. (\cite{loup93}). They are compared with astrometric data from Bergeat et al.
 (\cite{berge02a}, \cite{berge02b}=BKR) for (n) stars in common. Ratios to BKR means are then quoted in
 the last line (See text for a discussion).}  
\begin{flushleft}
{\scriptsize{
\begin{tabular}{cccccccccccccccccccc}
\hline\noalign{\smallskip}
n & O93 & BKR &   & n & G99 & BKR &  & n & O93 & SO1 & BKR &  & n & G02 & BKR &   & n & L93 & BKR\\
\noalign{\smallskip}\hline
\noalign{\smallskip}
104 & 707 & 708 &   & 11 & 1141 & 1161 &  & 54 & 567 & 427 & 614 &  & 25 & 1535 & 748 &  & 21 & 721 & 536\\
    & $\pm$ 218 & $\pm$ 258 &   &    & $\pm$ 459 & $\pm$ 452 &  &    & $\pm$ 171 & $\pm$ 166 & $\pm$ 218 &
&  & $\pm$ 927 & $\pm$ 357 &  &   & $\pm$ 424 & $\pm$ 182\\
    & 1.00 &    &    &    & 0.98 &  &   &    & 0.92 & 0.70 &  &   &    & 2.05 &    &   &   & 1.35\\
\noalign{\smallskip}\hline
\end{tabular}
}}
\label{mass}
\end{flushleft}
\end{table*}

   The predictions of Arndt et al. are compared to the data deduced from observations as 
   published by various authors. We use the effective temperatures of Bergeat et al.
   (\cite{bergea01}) as stellar temperatures. Using a new homogeneous scale constructed from
   model atmospheres, observed angular radii and spectral energy distributions, they have proved efficient
   in solving various problems. We also adopt the distances and luminosities derived by Bergeat et al. 
   (\cite{berge02b})
   from the HIPPARCOS data (\cite{ESA97}) by taking into account three distinct biases (their Sect. 2.3;
   see also Knapik et al. \cite{knapik98} and Bergeat et al. \cite{berge02a}). Pulsation masses were derived
   for carbon-rich LPVs by Bergeat et al. (\cite{berge02c}).
   The carbon to oxygen (C/O) ratios were taken from Lambert et al.
   (\cite{lamber86}), Olofsson et al. (\cite{olofs93b}) and Abia \& Isern (\cite{abia96}). The periods 
   of pulsation used are the photometric periods taken from the General Catalogue of Variable Stars (GCVS;
   Kholopov et al. \cite{GCVS}) or from the HIPPARCOS data (\cite{ESA97}). The values are those of 
   fundamental mode pulsators from the study of Bergeat et al. (\cite{berge02c}). Considering the 
   complexities of those pulsating atmospheres and the uncertainties on transfer in their models, we adopt
   the reference value $\rm{\Delta u\,=\,2\,km\,s^{-1}}$ of Arndt et al. (\cite{arndt97}) for the
   velocity amplitude of the pulsation. We shall see that its influence is quite limited.

   We re-scaled the mass loss rates from the literature to the distances of Bergeat et al. 
   (\cite{berge02b}) and a f-correction was applied to convert them to the scale of Sch\"{o}ier \& Olofsson 
   (\cite{schoie01}) who properly dealt with radiative transfer (Sect. 2). In addition,
   the expansion velocities and dust-to-gas density ratios were compiled for a sample 
   of carbon-rich giants (Sect. 2). The available predictions from hydrodynamical models are summarized 
   in Sect. 3 and equations fitted on a grid from the literature are adopted. Values of stellar 
   parameters from Bergeat et al. (\cite{bergea01}, \cite{berge02b}, \cite{berge02c}) were compiled for
   stars in common with millimetric studies and mean values collected for eight photometric groups 
   (Sect. 4) to be replaced in the formulae of Arndt et al. (\cite{arndt97}). Then, 
   the comparison of observed and predicted mass loss rates is performed in a diagram against effective
   temperature (Sect. 5). In a similar approach, the observed and predicted expansion velocities are
   compared by plotting {\it vs.} effective temperature (Sect. 6). The discrepancy noted between observed
   and predicted dust-to-gas density ratios, is estimated and discussed (Sect. 7). A few conclusions are
   given and discussed (Sect. 8) and simple approximate formulas are proposed in Appendix A.
\section{The data from millimetric observations}
   Mass loss rates, expansion velocities and dust-to-gas density ratios were extracted from the literature
   for a large sample of carbon-rich giants. The early millimetric data and results are summarized in a
   catalogue by Loup et al. (\cite{loup93}). Sch\"{o}ier \& Olofsson (\cite{schoie01}) revised a set of data
   from Olofsson et al. (\cite{olofs93a}, \cite{olofs93b}).
   The mass loss rate, for stars in common, is on average $3.1\pm 3.6$ times larger in the 2001 paper than 
   it was in those of 1993. Values in Groenewegen et al. (\cite{groen02a}) for 18 stars in common with Olofsson 
   et al. are on average 2 times larger. A mean 2.9 ratio was found from two stars in common with Winters 
   et al. (\cite{winter03}). Admittedly, recent accepted values for mass loss rates are typically 2-3 times 
   higher than the previous ones from Olofsson et al. (\cite{olofs93a}, \cite{olofs93b}). We also considered 
   data for a few stars from Groenewegen et al. (\cite{groene99}), Sch\"{o}ier \& Olofsson (\cite{schoie00}) 
   and Le Bertre et al. (\cite{lebert01}, \cite{lebert03}; see also Meixner et al. \cite{meixne98} for 
   RW LMi=CIT6 and Skinner et al. \cite{skinne98} for CW Leo=IRC+10216).
 \begin{table*}
 \caption
 {The data for 119 carbon-rich giants (Sect. 2) with variable star names (Columns 1 and 10, except for CGCS1006
 in Stephenson \cite{stephe89} and Alksnis et al. \cite{alksni01}) shown with boldface
 characters for 19 Miras. Data from SEDs and spectra are given: the photometric groups in Columns 2 and 11 
 with an additional J for $\rm{^{13}C}-$rich stars, the effective temperatures (Kelvin) in Columns 3 and 12,
 the carbon to oxygen abundance ratio (C/O) of the atmosphere in Columns 4 and 13, the absolute bolometric
 magnitude $\rm{M_{b}}$ in Columns 5 and 14, that can be translated into luminosity in solar units from 
 Eq. (3), and the distance $\rm{D_{BKR}}$ in parsec in Columns 6 and 15. 
 The data  from millimetric observations are displayed in Columns 7 and 16 (mass loss rate as 
 $\rm{\dot{M}'=10^{8}\,\dot{M}}$ where $\rm{\dot{M}}$ in solar mass per year),
 Columns 8 and 17 (expansion or terminal velocity $\rm{v_{e}}$ in $\rm{km\,s^{-1}},$ and Column 9 and 18
 ($\rm{r\,'=10^{3}\,r}$ where $\rm{r= \dot{M}_{d}/\dot{M}_{g}}$ is the dust-to-gas ratio of mass loss 
 rates). Additional notes deal with ranges: $\rm{^{a}}$ HC5-CV4 \& 3525-2775 K, $\rm{^{b}}$
 HC5-CV6 \& 3520-2645 K \& $\rm{M_{b}}= -3.72;-4.36,$ $\rm{^{c}}$ 2290-2245 K\&$\rm{M_{b}}= -4.86;-4.53,$ 
 $\rm{^{d}}$
 2920-2840 K \& $\rm{M_{b}}= -5.22;-4.80,$ $\rm{^{e}}$ $\rm{M_{b}}= -5.81;-5.48,$ $\rm{^{f}}$ 3345-2825 K
 \& $\rm{M_{b}}= -6.15;-5.67,$ $\rm{^{g}}$ 2405-2525 K \& $\rm{M_{b}}= -5.89;-6.10,$ $\rm{^{h}}$IRC+10216 
 1915-2105 K, $\rm{^{i}}$CIT6 2425-2465 K, $\rm{^{j}}$ 2100-1945 K \& $\rm{M_{b}}= -5.23;-4.60,$ $\rm{^{k}}$
 2865-2680 K \& $\rm{M_{b}}= -5.51;-5.00,$ $\rm{^{l}}$ CV5-CV6 \& 2650-2530 K \& $\rm{M_{b}}= -4.94;-5.06,$
 $\rm{^{m}}$ 1885-1875 K \& $\rm{M_{b}}= -6.28;-5.93,$ $\rm{^{n}}$ 2945-2865 K \& $\rm{M_{b}}= -4.84;-5.61,$
 $\rm{^{o}}$ 1975-1875 K, $\rm{^{p}}$ 2240-2095 K \& $\rm{M_{b}}= -5.42;-5.22,$ $\rm{^{q}}$ CV1-CV5 \&
 3245-2605 K \& $\rm{M_{b}}= -3.01;-3.69,$ $\rm{^{r}}$ 2735-2655 K \& $\rm{M_{b}}= -4.78;-4.58,$
 $\rm{^{s}}$ CIT 5, $\rm{^{t}}$ IRAS 05352+2247, $\rm{^{u}}$ 
 AFGL 971, $\rm{^{v}}$ AFGL 2067=IRC-10396, $\rm{^{w}}$ AFGL 3116=IRC+40540, $\rm{^{x}}$ AFGL 1235,
 $\rm{^{y}}$ AFGL 1961 (erroneously identified as V522Oph). Not shown in Fig.$~\ref{pmvste}:$
 according to Olofsson et al. (\cite{olofss00}), TT Cyg (which is surrounded by a thin shell) is presently
 losing mass at a modest rate of $\rm{3\,10^{-8}M_{\sun}\,yr^{-1}}$ which is perhaps to be multiplied by
 the 3.9 correction factor of Eq. (2); according to Izumiura et al. (\cite{izumiu96}), Y CVn** exhibited a
 $\rm{7-20\,10^{-6}M_{\sun}\,yr^{-1}}$ rate at the formation of its thin shell.}
\begin{flushleft}
{\scriptsize{
\begin{tabular}{cccccccccccccccccc}
\hline\noalign{\smallskip}
Name & G & $\rm{T_{eff}}$ & C/O & $\rm{M_{b}}$ & D & $\rm{\dot{M}'}$ & $\rm{v_{e}}$ & $\rm{r\,'}$ & 
Name & G & $\rm{T_{eff}}$ & C/O & $\rm{M_{b}}$ & D & $\rm{\dot{M}'}$ & $\rm{v_{e}}$ & $\rm{r\,'}$ \\
\noalign{\smallskip}\hline
\noalign{\smallskip}
VX And & CV6J & 2520 & 1.76 & -4.47 & 560 & 14 & 11.5 & 4.7 & 
AQ And & CV5 & 2660 & & -5.24 & 1015 & 43 & & \\
Z Psc & CV2 & 3095 & 1.014 & -4.65 & 465 & 6.1 & 3.5 & 2.9 & 
R Scl* & CV4 & 2625 & 1.34 & -5.48 & 475 & 640 & 16.9 & 0.86 \\
$\rm{\bf{R \,For}}$ & CV7 & 2000 &  & -5.82 & 865 & 500 & 16.5 & 0.87 & 
V623Cas & CV1J & 3360 &  & -4.49 & 515 & 11 &  &  \\
TW Hor & CV2 & 2950 &  & -5.28 & 485 & 20 & 5.5 & 2.5 & 
$\rm{\bf{V384Per}^{s}}$ & CV7 & 1820 &  & (-5.43) & 720 & 590 & 15.0 & 0.25\\
$\rm{\bf{Y \, Per}}$ & HC5$\rm{^{a}}$ & 3150$\rm{^{a}}$ &  & -4.09 & 975 & 29 & 7.2 & 0: & 
U Cam* & CV4 & 2695 & 1.30 & -5.09 & 525 & 200 & 20.6 & 1.0 \\
V466Per & CV4 & 2775 &  & (-4.66) & 530 & 24 & 9.0 & 2.3 & 
SY Per & CV5 & 2705 &  & -5.86 & 1430 & 150 & 17.5 & 0.79 \\
ST Cam & CV4 & 2805 & 1.14 & -6.07 & 800 & 110 & 9.0 & 0.89 & 
TT Tau & CV2 & 3090 &  & -4.88 & 585 & 20 & 5.0 & 0.42 \\
V346Aur & SCV & 2880 &  & -5.49 & 1110 & 20 &  &  & 
$\rm{\bf{AU \, Aur}}$ & CV5 & 2665 &  & -5.01 & 1470 & 57 &  &  \\ 
$\rm{\bf{R \, Ori}}$ & HC5$\rm{^{b}}$ & 3083$\rm{^{b}}$ &  & -3.72$\rm{^{b}}$ & 1700 & 61 & 10.0 & 0.67 & 
$\rm{\bf{R \, Lep}}$ & CV6 & 2290$\rm{^{c}}$ & 1.030  & -4.86$\rm{^{c}}$ & 335 & 160 & 18.0 & 1.4 \\ 
EL Aur & CV4 & 2730 &  & -4.60 & 650 & 28 &  &  & 
W Ori & CV5 & 2625 & 1.16 & -5.19 & 410 & 38 & 11.0 & 2.5 \\ 
V431Ori & CV6 & 2540 &  & -3.78 & 540 & 22 &  &  & 
$\rm{\bf{UV \, Aur}}$ & CV3 & 2920$\rm{^{d}}$ &  & -5.22$\rm{^{d}}$ & 1090 & 160 & 10: & \\ 
S Aur & CV7 & 1940 & 1.014 & (-5.43) & 1130 & 420 & 25.5 & 1.0 & 
RT Ori & CV4 & 2870 &  & -4.31 & 675 & 22 &  &  \\ 
CGCS1006$\rm{^{t}}$ & CV6 & 2300 &  & (-5.05) & 1870 & 43 &  &  & 
TU Tau & CV3 & 2850 &  & -5.81$\rm{^{e}}$ & 1055 & 57 &  &  \\ 
Y Tau & CV4 & 2735 & 1.040 & -6.00 & 735 & 160 & 11.0 & 1.6 & 
W Pic & CV6 & 2530 &  & -4.85 & 665 & 51 & 16.0 & 1.7 \\ 
FU Aur & CV2 & 3035 &  & -5.56 & 1295 & 28 &  &  & 
TU Gem & CV4 & 2715 & 1.12 & (-4.66) & 515 & 44 & 11.5 & 1.7 \\ 
FU Mon & SCV & 3085$\rm{^{f}}$ &  & -6.15$\rm{^{f}}$ & 1450 & 26 &  &  & 
$\rm{\bf{V \, Aur}}$ & CV4 & 2820 &  & -5.47 & 1760 & 360 & 20 & 0.13 \\ 
BN Mon & CV6 & 2410 &  & (-5.05) & 1280 & 420 & 24.2 & 0.73 & 
ZZ Gem & CV6 & 2530 &  & (-5.05) & 1835 & 59 & 6.9 & 0.22 \\ 
BL Ori & CV2 & 3035 & 1.039 & -5.13 & 595 & 25 & 9.0 & 1.2 & 
CR Gem & CV2 & 2960 &  & -5.60 & 920 & 84 & 14.9 & 1.3 \\ 
UU Aur & CV4 & 2760 & 1.063 & -5.78 & 415 & 76 & 11.0 & 0.16 & 
NP Pup & CV2 & 3090 &  & -4.54 & 510 & 19 & 9.5 & 1.0 \\ 
RV Mon & CV3 & 2910 &  & -4.90 & 670 & 17 &  &  & 
V614Mon & CV1 & 3320 &  & -3.47 & 485 & 1.5 &  &  \\ 
RY Mon & CV6 & 2440 & 1.19 & -4.90 & 685 & 59 & 11.0 & 1.1 & 
W CMa & CV3 & 2975 & 1.046 & -5.42 & 785 & 61 & 10.5 & 0.63 \\ 
V406Pup & CV3 & 2875 &  & -4.15 & 610 & 11 &  &  & 
RU Pup & CV3 & 2680 &  & -3.33 & 455 & 13 &  &  \\ 
$\rm{\bf{V346Pup}^{x}}$ & CV7 & 1875 &  & (-5.43) & 1120 & 3310 & 20.7 & 0.50 & 
R Pyx & CV6 & 2440 &  & -4.52 & 1175 & 68 & 8.8 & 1.2 \\
UZ Pyx & CV1 & 3325 & 1.30 & -4.63 & 815 & 26 &  &  & 
X Cnc & CV5 & 2645 & 1.14 & -5.77 & 710 & 62 & 7.0 & 1.3 \\ 
T Cnc & CV6 & 2405$\rm{^{g}}$ &  & -5.89$\rm{^{g}}$ & 1085 & 97 &  &  & 
$\rm{\bf{CW \, Leo \, ^{h}}}$ & CV7 & 1915$\rm{^{h}}$ & 1.4 & -5.83 & 150 & 3300 & 14.5 & 0.66 \\ 
Y Hya & CV5 & 2645 & 1.52 & -4.89 & 485 & 37 & 9.0 &  & 
X Vel & CV5 & 2700 & 1.16 & -5.46 & 620 & 63 & 10.0 & 0.68 \\
SZ Car & CV4 & 2810 &  & (-4.66) & 365 & 46 & 14.0 & 0.30 & 
RW LMi$\rm{^{i}}$ & CV6 & 2445$\rm{^{i}}$ &  & (-5.05) & 410 & 650 & 17.0 & 4.5 \\ 
XZ Vel & CV6 & 2430 & 1.22 & (-5.05) & 760 & 94 & 14.0 & 0.90 & 
$\rm{\bf{CZ \, Hya}}$ & CV5 & 2525 & 1.02 & (-4.82) & 990 & 125 & 12.0 & 0.65 \\ 
U Ant* & CV4 & 2810 & 1.44 & -4.93 & 320 & 200 & 21.8 &  & 
U Hya & CV3 & 2965 & 1.043 & -3.93 & 175 & 21 & 7.0 & 0.98 \\ 
VY UMa & CV2 & 2930 & 1.060 & -4.67 & 445 & 14 & 6.0 & 0.97 & 
V Hya & CV6 & 2160 & 1.050 & -5.88 & 495 & 750 & 24.2 & 1.1 \\ 
RR Mus & CV2 & 3090 & 1.010 & -4.34 & 690 & 27 &  &  & 
SS Vir & CV6 & 2560 & 1.080 & (-5.05) & 560 & 36 & 12.5 & 1.4 \\ 
Y CVn** & CV5J & 2760 & 1.087 & -4.64 & 260 & 14 & 8.5 & 2.0 & 
DY Dru & CV6 & 2420 & 1.150 & (-5.05) & 900 & 47 &  &  \\ 
$\rm{\bf{RU \, Vir}}$ & CV6 & 2100$\rm{^{j}}$ &  & -5.23$\rm{^{j}}$ & 675 & 230 & 18.4 &  & 
RX Cru & CV6 & 2650 &  & (-5.05) & 845 & 62 &  &  \\ 
RY Dra & CV5J & 2810 & 1.18 & -5.28 & 550 & 44 & 10.0 & 0.27 & 
$\rm{\bf{RV \, Cen}}$ & CV3 & 2865$\rm{^{k}}$ & 1.030 & -5.51$\rm{^{k}}$ & 940 & 78 & 12.6 & 1.3 \\ 
NSV6507 & CV3 & 2920 &  & (-4.39) & 560 & 18 & 6.5 & 3.1 & 
V996Cen & CV4 & 2695 & 1.53 & -5.81 & 830 & 44 & 11.0 & 1.1 \\ 
Z Lup & CV5 & 2655 & 1.12 & -5.05 & 1105 & 20 &  &  & 
X TrA & CV5 & 2710 & 1.17 & -5.71 & 460 & 18 & 9.1 & 0.81 \\ 
AS Cir & CV6 & 2420 &  & (-5.05) & 935 & 78 &  &  & 
U Aps & CV5 & 2695 &  & -4.68 & 825 & 28 &  &  \\ 
$\rm{\bf{V \, CrB}}$ & CV7 & 2090 &  & -4.80 & 845 & 130 & 7.5 & 1.3 &
$\rm{\bf{V \, Oph}}$ & CV3 & 3010 &  & -4.25 & 580 & 14 & 7.8 & 0.64 \\
SU Sco & CV5 & 2655 &  & (-4.82) & 635 & 22 &  &  & 
V1079Sco & CV6 & 2510 & 1.10 & -5.33 & 920 & 120 &  &  \\ 
V2309Oph$\rm{^{y}}$ & CV6 & 2425 & 1.10 & (-5.05) & 1025 & 34 &  &  & 
TW Oph & CV6 & 2440 & 1.20 & -5.12 & 495 & 23 & 8.6 & 0.15 \\
TT Sco & CV6 & 2430 &  & (-5.05) & 865 & 57 &  &  & 
V Pav & CV6 & 2545 &  & -4.74 & 430 & 66 & 16.0 & 1.1 \\ 
SX Sco & CV4 & 2785 &  & -5.02 & 830 & 38 &  &  & 
$\rm{\bf{T \, Dra}}$ & CV6 & 1850 &  & -5.83 & 1425 & 820 & 13.5 & 1.3 \\ 
ES Ser & CV6 & 2500 &  & (-5.05) & 905 & 30 &  &  & 
TY Oph & CV5 & 2680 &  & (-4.82) & 845 & 32 &  &  \\
T Lyr & CV6J & 2310 & 1.29 & -5.43 & 645 & 80 & 11.5 & 2.2 &
HK Lyr & CV3 & 2945 &  & -4.66 & 730 & 26 &  &  \\ 
HK Lyr & CV5 & 2620 &  & -4.39 & 730 & 26 &  &  & 
DR Ser & CV5 & 2650 &  & -5.40 & 1295 & 100 & 20.1 & 3.9 \\
S Sct* & CV4 & 2755 & 1.069 & -5.18 & 580 & 560 & 17.3 &  &
UV Aql & CV5 & 2700 & 1.11 & -5.01 & 910 & 30 &  & \\
V Aql & CV6 & 2525 & 1.25 & -5.65 & 560 & 66 & 8.5 & 1.0 & 
CG Vul & CV5 & 2685 &  & (-4.82) & 805 & 33 &  &  \\ 
V1942Sgr & CV2 & 2960 & 1.12 & -4.60 & 550 & 19 & 10.0 & 1.1 & 
UX Dra & CV2 & 3090 & 1.046 & -5.53 & 545 & 37 & 4.0 & 0.49 \\ 
AQ Sgr & CV4 & 2790 & 1.033 & -5.66 & 790 & 77 & 10.0 & 0.63 & 
V391Aql & CV6 & 2585 &  & -5.05 & 1950 & 38 &  & \\ 
TT Cyg* & CV4 & 2825 &  & -3.98 & 620 & (12:) & 13.5 &  & 
AX Cyg & CV5 & 2655 &  & -4.03 & 520 & 23 &  &  \\ 
X Sge & CV5 & 2630 &  & -4.60 & 860 & 24 &  &  &
SV Cyg & CV5 & 2600 &  & -4.22 & 700 & 19 &  &  \\ 
RS Cyg & CV2 & 3100 &  & -5.08 & 655 & 20 &  &  &
RT Cap & CV6 & 2480 & 1.10 & -4.95 & 560 & 23 & 8.0 & 1.8 \\
$\rm{\bf{U \, Cyg}}$ & CV5$\rm{^{l}}$ & 2650$\rm{^{l}}$ &  & -4.94$\rm{^{l}}$ & 700 & 100 & 13.0 & 1.5 &
$\rm{\bf{V \, Cyg}}$ & CV7 & 1880$\rm{^{m}}$ &  & -6.28$\rm{^{m}}$ & 740 & 630 & 11.5 & 0.72 \\
T Ind & CV2 & 2990 &  & -5.36 & 645 & 17 & 6.0 & 2.0 &
Y Pav & CV3 & 2945$\rm{^{n}}$ &  & -4.84$\rm{^{n}}$ & 420 & 23 & 8.0 & 1.4 \\
$\rm{\bf{V1426Cyg}}$ & CV7 & 1975$\rm{^{o}}$ &  & (-5.43) & 820 & 540 & 14.0 & 1.3 &
$\rm{\bf{S \, Cep}}$ & CV6 & 2240$\rm{^{p}}$ &  & -5.42$\rm{^{p}}$ & 495 & 290 & 22.0 & 1.2 \\
V460Cyg & CV2 & 2950 & 1.062 & -5.81 & 635 & 45 & 10.0 & 0.78 &
RV Cyg & CV5 & 2675 & 1.20 & -5.57 & 640 & 140 & 13.5 & 1.2 \\
LW Cyg & CV5 & 2580 &  & -5.45 & 1095 & 35 &  &  &
$\rm{\bf{RZ \, Peg}}$ & CV3$\rm{^{q}}$ & 2925$\rm{^{q}}$ &  & -3.3$\rm{^{q}}$ & 605 & 10 & 12.6 & 0.59 \\
DG Cep & CV2 & 2985 &  & -4.71 & 800 & 23 &  &  &
TX Psc & CV2 & 3125 & 1.027 & -5.09 & 315 & 32 & 10.7 & 0.72 \\
WZ Cas & CV2 & 3095 & 1.010 & -5.48 & 650 & 13 & 3.8 & 1.0 &
V688Mon$\rm{^{u}}$ & CV7 & 1670 &  & (-5.43) & 1370 & 2400 & 13.4 & 4.1 \\
FX Ser$\rm{^{v}}$ & CV7 & 2050 &  & (-5.43) & 1230 & 3100 & 28.4 & 0.58 &
$\rm{\bf{LP \, And}}\rm{^{w}}$ & CV7 & 2040 &  & (-5.43) & 840 & 2100 & 14.0 & 1.7 \\
\noalign{\smallskip}\hline
\end{tabular}
}}
\label{mass}
\end{flushleft}
\end{table*}

   Usually, the distances adopted by the authors substantially differ and this is a fundamental problem
   since the mass loss rate (for both gas and dust) increases as the squared distance.
   Large discrepancies may exist between distances estimated from various methods: mean absolute magnitude
   adopted, say in the K-band, mean bolometric magnitude with bolometric correction or mean luminosity,
   absolute magnitude {\it vs. \/} period for long period variables, kinematical methods ... The 
   astrometric parallaxes of those bright distant giants are also affected by substantial errors. First of 
   all, we calculated the mean distances and dispersions for the various samples of the above-mentioned
   references and compared them to their counterparts from Bergeat et al. (\cite{berge02b}, BKR) whose
   values are astrometric in essence and deduced from HIPPARCOS data (Knapik et al. \cite{knapik98} and
   Bergeat et al. \cite{berge02a}), with no photometric criteria used. The results are shown in 
   Table 1. There is apparently an excellent agreement for the 104 stars of Olofsson et al. 
   (\cite{olofs93a}) in common, with 707-708 pc as mean values. A detailed analysis (Bergeat \cite{bergea04})
   however proved this is fortuitous since a Malmquist-type bias occurred in the Olofsson et al. distances
   in this sample of magnitudes $\rm{K \le 2},$ adopting $\rm{M_{K}}\simeq -8.1$ (actually there is a
   3-magnitude range whose central value is close to -7.50). Conversely, no such bias appeared when
   comparing the distances deduced from the $\rm{M_{K}}-$magnitudes derived by Mennessier et al. 
   (\cite{mennes01}) through a maximum likelihood method for 124 carbon giants in common with BKR. 
   There is also a good agreement with BKR in Table  1 for a small sample of 11 stars from Groenewegen et al. 
   (\cite{groene99}). They are located farther away at 1141-1161 pc on average, that is 1.63 times more
   distant. We also considered a sample of 54 stars in common to Olofsson et al. (\cite{olofs93a})
   and Sch\"{o}ier \& Olofsson (\cite{schoie01}). There is a 1.33 ratio (i.e. 1.76 on
   mass loss rates), in the direction opposite to that of the 3.1 mean ratio of mass loss rates mentioned
   above. It is to be compared to 0.92 when referred to BKR. The Groenewegen et al. (\cite{groen02b}) mean
   distance is 2.05 times larger than the BKR-value for 25 stars in common, leading to their 
   mass loss rates being systematically larger. The factor is only 1.35 between Loup et al. (\cite{loup93}) 
   and BKR.

   Finally, for the sake of consistency, we transformed the rates $\rm{\dot{M}}$ given at distances D in 
   the mentioned papers, according to the relation
   \begin{equation}
    \rm{\dot{M}_{BKR}\:=\:\dot{M}\:\left(D_{BKR}\,/\,D\right)^{2}}
   \end{equation}
   where $\rm{D_{BKR}}$ stands for the distances of Bergeat et al. (\cite{berge02b}). 
   The gas mass loss rates were derived from CO observations by a formula established by
   Knapp \& Morris (\cite{knapp85}) for unresolved optically thick envelopes. From a more detailed
   radiative transfer analysis, Sch\"{o}ier \& Olofsson (\cite{schoie01}) found that the rates resulting
   from the formula are systematically underestimated when compared to the new values (their Fig. 8).
   As expected, the discrepancy increases with decreasing rate i.e. decreasing optical depth. From the
   comparison of $\rm{\dot{M}_{BKR}}$ for 53 stars in common we obtained a mean correction factor (f) of
   $2.5\pm 1.7$ to scale the rates from Olofsson et al. (\cite{olofs93a}) to the Sch\"{o}ier \& Olofsson 
   (\cite{schoie01}) level. We adopted mean values in four ranges
   \begin{eqnarray}
   0.94\pm 0.34 \simeq 1 \;\; \rm{if}\;\; \rm{\dot{M}_{BKR}}>4\,10^{-6} \nonumber \\
   2.2\pm 1.0 \;\; \rm{if} \;\; 4\,10^{-7}<\rm{\dot{M}_{BKR}}<4\,10^{-6} \nonumber \\
   2.5\pm 1.4 \;\; \rm{if} \;\; 7\,10^{-8}<\rm{\dot{M}_{BKR}}<4\,10^{-7} \nonumber \\
   3.9\pm 3.1 \;\; \rm{if} \;\; \rm{\dot{M}_{BKR}}<7\,10^{-8} 
   \end{eqnarray}
   where quoted rates refer to values from Olofsson et al. (\cite{olofs93a}). This is in fair agreement
   with Fig. 8 in Sch\"{o}ier \& Olofsson (\cite{schoie01}). About 70\% of the data we used were taken 
   from those two references. From 17 stars in common, we found $\rm{f=1.9\pm1.4}$ for Groenewegen et al.
   (\cite{groen02b}, 8\%) while $\rm{f=3.3\pm0.3}$ was deduced from the seven values of Groenewegen et al.
   (\cite{groene99}, 2.7\%). Here too, there is some indication of f tending to unity for objects with very
   large rates but we were unable to split the samples. For Loup et al. (\cite{loup93}, 7 stars and 9\%), 
   we obtained $\rm{f=1.2\pm0.5}$ while $\rm{f=3.4\pm2.1}$ and $\rm{f=3.1\pm1.7}$ were deduced for
   Le Bertre et al. (\cite{lebert01}, 5 stars) and (\cite{lebert03}, 4 stars) respectively. 
   These later five correction factors and those in Eq. (2) were applied to the $\rm{\dot{M}_{BKR}}-$data 
   derived from Eq. (1). We rejected the data available in Winters et al. (\cite{winter03}) since the 
   correction factor curiously amounted to 0.4 from only two stars in common with Sch\"{o}ier \& Olofsson.
   
   Both the obtained mean gas mass loss rates and the $\rm{D_{BKR}}-$distances are given in Table 2 
   together with photometric groups and effective temperatures from Bergeat et al. (\cite{bergea01}). 
   The luminosities in solar units can be derived from the absolute bolometric magnitudes of Bergeat et 
   al. (\cite{berge02b}) making use of
   \begin{equation}
   \rm{L/L_{\sun}\:=\:10^{-0.4\,\left(M_{b}\,-\,4.73\right)}} \:\:\:\:\:.
   \end{equation} 
   Additional star identifiers and detailed bibliographic sources of the mass loss data for each star are 
   provided in Table 3, only available in electronic form. The C/O ratios quoted in Table 2 were 
   taken from Lambert et al. (\cite{lamber86}), Olofsson et al. (\cite{olofs93b}) and Abia \& Isern 
   (\cite{abia96}). Also given are the expansion velocities from the same five references mentioned above
   for mass loss rates. Both the dust and gas mass loss rates are proportional to the adopted squared 
   distance and no corresponding corrections need be applied to the dust-to-gas ratios. The dust mass
   loss rate is derived from the IRAS (60$\rm{\mu m}$) making use of a formula given by Sopka et al.
   (\cite{sopka85}; see e.g. Eq. (8) in Olofsson et al. (\cite{olofs93a}). The dust rate is 
   proportional to $\rm{L^{-0.5}}.$ We thus scaled the values from Olofsson et al. (\cite{olofs93b}),
   Groenewegen et al. (\cite{groene99}) and Groenewegen et al. (\cite{groen02b}) to the luminosities
   quoted in Table 2, according to the individual luminosities adopted by these authors. The obtained
   values were then divided by the correction factors of Eq. (2) or by 3.3 or 1.9, i.e. the f-factors 
   for the gas mass loss rates from those three references. As discussed in Sect. 7, the latter correction 
   strongly shifts observations toward model predictions. Two estimates are provided by Olofsson et al. 
   (\cite{olofs93a}), namely a lower one assuming no drift velocity for the dust, and an upper one adopting
   an upper limit for drift velocities (up to 20 km/s, i.e. eventually higher than the gas expansion 
   velocity) obtained by equating the radiation force on the grains and the drag force due to gas-grain 
   collisions in the limit of supersonic motion (e.g. Kwok \cite{kwok75}). The estimates from
   Groenewegen et al. make use of moderate drift velocities, usually 2-3 km/s. Their mean values, 
   $\left(1.2\pm1.3\right)\,10^{-3}$ for Groenewegen et al. (\cite{groene99}, n=8) and 
   $\left(1.45\pm1.1\right)\,10^{-3}$ for Groenewegen et al. (\cite{groen02b}, n=24)
   are consistent with that of the ``lower'' values ($\left(1.3\pm1.0\right)\,10^{-3}$, n=55) of Olofsson 
   et al. (\cite{olofs93a}), while the ``upper'' values yield a significantly larger mean 
   ($\left(3.2\pm2.5\right)\,10^{-3}$, n=55). Average values of the available data 
   are quoted in Table 2, ignoring the ``upper'' ones in Olofsson et al. (\cite{olofs93a}). They thus
   correspond to the low drift velocities adopted here.
   
   The sample of Table 2 has 119 carbon-rich giants with at least one of the three
   quantities from millimetric data given. Many ``optical'' carbon stars of low to moderate mass loss 
   rates are included while the ``infrared'' carbon stars with higher rates and often optically thick
   dust shells are fewer. This is due to our requirement that stellar data be available from Bergeat et
   al. (\cite{bergea01}, \cite{berge02b}, \cite{berge02c}) that, in turn, necessitated sufficient
   photometric data in the visible range. This is very often missing for ``infrared'' carbon stars. This 
   is rather unfortunate since 330 ``infrared'' carbon stars were studied by Groenewegen et al. 
   (\cite{groen02a}, \cite{groen02b}).
\section{The predictions from hydrodynamical models}
   As mentioned in Sect. 1, both pulsation and dust formation are involved in the mass loss phenomenon on 
   the asymptotic branch. The mass loss relation of Dominik et al. (\cite{domini90}) is based on calculated 
   models and was the first one available. It describes the stationary outflows of carbon-rich low mass 
   stars. {\it The mass loss rate increases with increasing luminosity and decreasing effective 
   temperature.\/} The strong dependence upon the effective temperature is due to the fact that dust always
   condenses at about the same temperature. For stars with higher effective temperatures, this
   condensation temperature is reached farther out in the shell where dust formation is less effective.
   {\it The mass loss rate increases as the stellar mass decreases.\/} Other things being equal,
   increased gravity due to larger masses inhibits the wind. The dependence of the mass loss rate on the
   carbon to oxygen ratio is much less marked, especially for values larger than 1.5. Dominik et al.
   proposed an analytical fit, their formula (7), which is presumably accurate for extreme mass loss rates,
   i.e. for $\rm{\dot{M}\ge 10^{-6}\,M_{\sun}yr^{-1}}.$ The terminal velocity ranges from 5 to 
   40$\rm{\,km\,s^{-1}}.$ {\it It increases with decreasing effective temperature and with increasing 
   luminosity or increasing C/O ratio.\/} The authors proposed a linear fit as their Eq. (8). The
   dust-to-gas density ratio is closely related to the final degree of condensation, as shown by formula 
   (9) of Dominik et al., and it ranges from $6\,10^{-4}$ to $5\,10^{-3}.$

   Making use of the models of Fleischer et al. (\cite{fleisc92}) and Fleischer (\cite{fleisc94}), Arndt
   et al. (\cite{arndt97}) derived a grid of 48 dynamical models usually with less extreme mass loss than
   in the cases explored by Dominik et al. (\cite{domini90}). They are realistic models of circumstellar
   dust shells around carbon-rich long-period variables. They include the calculation of time-dependent 
   hydrodynamics as well as a detailed treatment of dust formation, growth and evaporation. The maximum 
   likelihood method was applied to the grid with six parameters they considered as independent,
   namely the stellar temperature, the stellar luminosity, the stellar mass, the abundance ratio of carbon
   to oxygen, the pulsational period and the velocity amplitude of the pulsation. They obtained their
   equations (1), (3) and (5) for the mass loss rate, the dust-to-gas ratio and the expansion (outflow) 
   velocity, respectively. The trends are essentially the same as those obtained by Dominik et al. and
   summarized above. Keeping constant the parameters of low influence, they also wrote the simplified
   relations they numbered (2), (4) and (6) for the same three ``mass loss'' quantities.

   Applying a piston and a variable luminosity at the inner boundary in the models of H\"{o}fner et al. 
   (\cite{hofner95}, \cite{hofner96}), H\"{o}fner \& Dorfi (\cite{hofner97}) constructed a set of 22
   carbon-rich LPV models. The ranges of the parameters were smaller but Arndt et al. (\cite{arndt97})
   conclude a good agreement with their work for values in common.

   We recall here the equations (1), (3) and (5) from Arndt et al. (\cite{arndt97}) whose predictions we
   intend to compare with the observed data as collected in Sect. 2. For the mass loss rate in solar mass 
   per year they derived
   \begin{eqnarray}
    \rm{\log \dot{M}=-4.95-9.45\log \left(T/2600\right)+1.65\log \left(L/10^{4}L_{\sun}\right)} \nonumber\\
    \rm{-2.86\log \left(M/M_{\sun}\right)+0.47\log \left(\epsilon _{C}/1.8\epsilon _{O}\right)} \nonumber\\
    \rm{-0.146\log \left(P/650\right)+0.449\log \left(\Delta u/2.0\right)} \:\:\:\:\: 
   \end{eqnarray}
   where the parameters of major influence are the effective temperature T in Kelvin, the luminosity L and 
   the mass M.
   The minor parameters are the carbon to oxygen abundance ratio, the period of the LPV in
   days, and the piston velocity amplitude in $\rm{km\,s^{-1}}.$ For the dust-to-gas density ratio, they 
   obtained
   \begin{eqnarray}
    \rm{\log \left(\rho _{d}/\rho _{g}\right)=2.74-1.62\log T-0.248\log \left(L/L_{\sun}\right)}\nonumber\\
    \rm{-0.658\log \left(M/M_{\sun}\right)+2.45\log \left(\epsilon _{C}/\epsilon _{O}\right)}\nonumber\\
    \rm{+0.230\log P-4.93\,10^{-3}\log \left(\Delta u\right)} \:\:\:\:\:
   \end{eqnarray} 
   with the same units. In addition to T, L and M, the carbon to oxygen abundance ratio is here a major 
   parameter. Again, the period and piston velocity amplitude have little influence. Finally, they
   give for the expansion (or outflow) velocity
   \begin{eqnarray}
    \rm{\log v _{\infty} =0.730-0.148\log T+0.107\log \left(L/L_{\sun}\right)}\nonumber\\
    \rm{-6.40\,10^{-2}\log \left(M/M_{\sun}\right)+1.74\log \left(\epsilon _{C}/\epsilon _{O}\right)}\nonumber\\
    \rm{+9.56\,10^{-2}\log P-7.07\,10^{-4}\log \left(\Delta u\right)} \:\:\:\:\:
   \end{eqnarray}
   where the effective temperature, the luminosity and the carbon to oxygen abundance ratio prove to be the
   main parameters. The stellar mass, the period of pulsation and the piston velocity amplitude have little
   influence here. 
\section{The realm of carbon-rich AGB stars}
   The adopted values of the six parameters in the models mentioned in Sect. 3 were varied more or less
   freely by the authors. The resulting Eqs. (4), (5) and (6) are given for ranges of parameter
   values that are those of the used (satisfactory) models.
   {\it Each parameter is supposed to vary independently which is actually not the case for
   existing carbon-rich giants on the AGB.\/} We briefly summarize here the results from Bergeat et al.
   (\cite{bergea01}, \cite{berge02b}, \cite{berge02c}) and references therein, on this specific topic. The
   bright carbon-rich giants
   are located on the asymptotic branch, with various initial masses and chemical compositions. Consequently,
   those Galactic giants populate a strip in the HR diagram with increasing luminosity for
   decreasing effective temperature (See Figs. 8 and 9 of Bergeat et al. \cite{berge02b}). Mean values of
   both quantities were given in their Table 3 for the HC-CV groups. This classification was shown to be
   strongly correlated with effective temperature decreasing along the sequence HC0 to HC5 and then CV1 to
   CV7 (Bergeat et al. \cite{bergea01}). Influenced by true dispersion and errors on distance estimates, the
   populated
   strip is rather large but the trend in mean values is obvious. A study of pulsation modes resulted in
   the evaluation of pulsation masses in good agreement with evolution tracks in the HR diagram (Bergeat et
   al. \cite{berge02c}) for long period variables. Mean pulsation masses and periods increase with
   decreasing effective temperature along the sequence of photometric groups, as shown in their Table 1.
   In a mass-luminosity diagram (their Fig. 4), the mean luminosity increases with increasing mean mass.
   For the cool carbon variables (CV-stars), the carbon to oxygen abundance ratio was plotted against effective
   temperature in Fig. 7 of Bergeat et al. (\cite{berge02c}). The mean values and dispersions are increasing
   with decreasing effective temperatures along the CV1 to CV6 sequence. They however decrease below 2500 K
   at the CV6-CV7 junction, a result possibly due to large dust condensation (SiC and carbonaceous grains)
   in the atmosphere and resulting carbon depletion in the gas phase. The only parameter of formulae
   (4) to (6) in Sect. 3 for which we
   have no indication, is the piston amplitude velocity. It is however supposed to be higher in Miras and
   semi-regulars with large amplitudes than it is in semi-regulars with low amplitudes and irregular variables.
   Its influence is however small on the three quantities calculated from Eqs. (4) to (6). We thus adopted
   the $\rm{\Delta u=2\,km\,s^{-1}}$ value as favored by Arndt et al. (\cite{arndt97}). We have summarized
   the values in Columns 7 to 9 of Table 4 as calculated from the mean data of Columns 2 to 6, following
   the same sequence of photometric groups. In those calculations, we eventually went beyond the validity
   ranges of formulae (4) to (6) {\it stricto sensu}, as defined by the parameter values that Arndt et al.
   introduced in their 48 models. Discussions are postponed to the following sections.
\section{Mass loss rate {\it vs.} effective temperature}
\begin{figure*}
    \centering
    \resizebox{14.0cm}{!}{\includegraphics{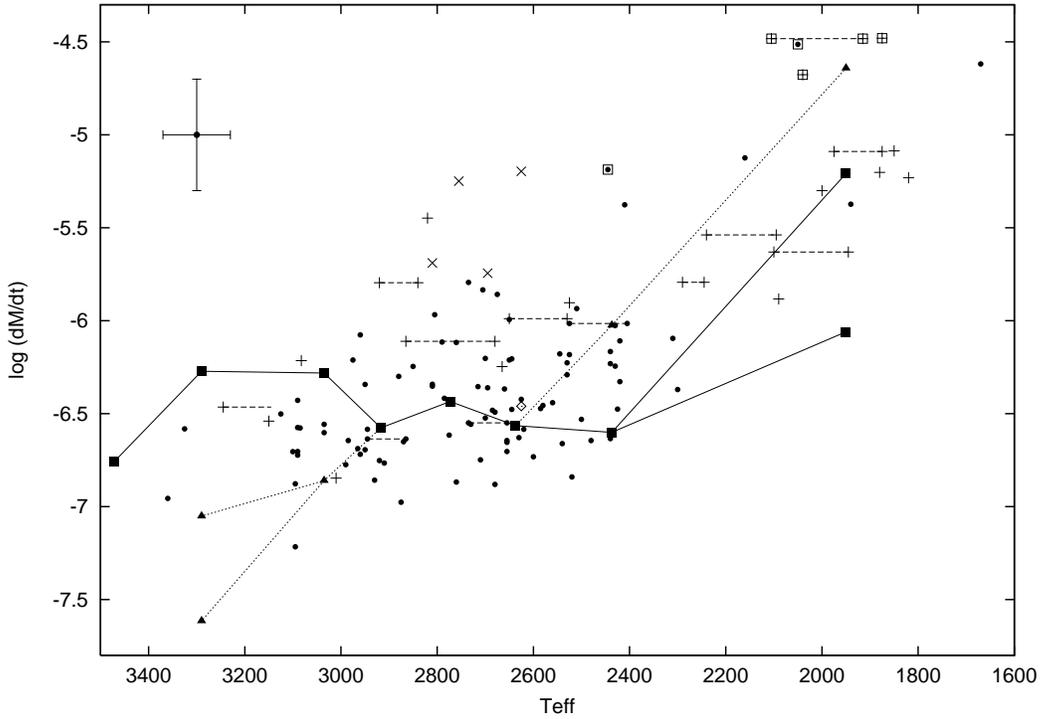}}
    \caption{Logarithm of the mass loss rate in solar mass per year as a function of effective temperature
     in Kelvin, as taken from Table 2 (observations). Typical error bars are displayed at left.
     Semi-regulars
     and Miras are shown as filled circles and crosses respectively with 4 non-Miras with detached shells
     ($\times$ and diamond symbol for present R Scl), and 5 objects with optically-thick dust shells
     (squared symbols). For a few variables, there are two symbols for temperatures obtained at distinct
     phases. The predictions from models (Table 4) are plotted as filled
     squares connected by full lines. Additional predictions for various mean masses are shown as
     filled triangles connected with dashed lines (see text).}
    \label{pmvste}
    \end{figure*}

   The mass loss rates $\rm{\dot{M}}$ as quoted in Table 2 (Columns 7 and 16) are plotted on a logarithmic
   scale in Fig.$~\ref{pmvste}$ as a function of the effective temperatures from Columns 3 and 12.
   The main trend of increasing mass loss rate with decreasing effective temperature is observed as
   expected from theory (Sect. 3).
   Typical error bars on both coordinates are also shown. The Miras are displayed as
   crosses while the filled circles stand for semi-regulars. Among the latter, four stars with detached
   shell(s) are shown with $\times$-symbols in the 2600-2800 K range. Denoted by asterisks in Table 2,
   they are U Ant, S Sct, R Scl and U Cam. The observation of detached shell(s) is indicative of recent
   enhanced mass loss. As a consequence, those four stars do exhibit larger mass loss rates than the
   bulk of our sample, in the same temperature range. Recent observations of a HCN line emission in the
   central source of R Scl (2625 K) that was resolved were secured with a FWHM of 1" by Wong et al. 
   (\cite{wong04}). The
   authors thus derive for the {\it present\/} mass loss rate $\rm{2\,10^{-7}\,M_{\sun}yr^{-1}}$ for an
   adopted distance of 360 pc, a rate we convert into $\rm{3.5\,10^{-7}\,M_{\sun}yr^{-1}}$ for our 475 pc
   distance estimate (see Column 15 of Table 2). This latter value, displayed as a diamond-symbol in
   Fig.$~\ref{pmvste},$ is located in the main locus. This is about 16 times less than the earlier
   $\rm{5.5\,10^{-6}\,M_{\sun}yr^{-1}}$ value quoted for R Scl in Column 16 of Table 2, a rather drastic
   change. This is convincing evidence for highly episodic mass loss, possibly caused by a thermal pulse.
   The detached shell of R Scl has recently been investigated for MgS, mining the mass-loss history of this
   object (Hony \& Bouwman \cite{hony04}; see also Olofsson et al. \cite{olofss00} for a detailed study
   of circumstellar CO in TT Cyg). We conclude that mass loss rates larger than about 
   $\rm{10^{-6}\,M_{\sun}yr^{-1}}$ in the 2500-2800 K range are representative of temporarily-enhanced mass
   loss leading to detached shell(s) hundreds or a few thousands of years later (see e.g. Lindqvist et al
   \cite{lindqv96} for U Cam, Izumiura et al. \cite{izumiu97} for U Ant, Izumiura et al. \cite{izumiu96}
   and Le Bertre \& G\'erard \cite{lebert04} for Y CVn).

   At lower temperatures, 
   we used symbols surrounded by squares for five stars with optically-thick dust shells as shown
   by large infrared excesses. They are CW Leo (IRC+10216), V 346 Pup and LP And (three Miras),
   and RW LMi (CIT 6) and FX Ser (two semi-regulars). They exhibit mass loss rates much larger than 
   those from the carbon-rich giants with optically-thin dust shells in the same 2500-1800 K range, that
   is larger than about $\rm{6\,10^{-6}\,M_{\sun}yr^{-1}}$ for effective temperatures less than 2500 K.
   A few intermediate objects are also noticed (V Hya, BN Mon).
\setcounter{table}{3}
 \begin{table}
 \caption
 {The mean stellar data for carbon-rich giants (Columns 2-6) from Bergeat et al. (\cite{bergea01}, 
 \cite{berge02b}, \cite{berge02c}) and the corresponding mass loss data (Columns 7-9) calculated from
 formulae (4), (5) and (6) taken from Arndt et al. (\cite{arndt97}). For photometric groups of Column 1,
 the mean values of effective temperatures in Kelvin (Column 2), the mean luminosities $\rm{l=<L/L_{\sun}>}$
 (Column 3) and masses $\rm{m=<M/M_{\sun}>}$ (Column 4) in solar units, the mean carbon to oxygen abundance
 ratio C/O (Column 5) and the mean pulsation period P in days (Column 6).
 The group CV5 ``und'' corresponds to a small sample of underluminous CV5-stars which are not included. 
 Adopting the piston velocity amplitude $\rm{\Delta u=2\,km\,s^{-1}}$ favored by Arndt et al., the
 corresponding mass loss rates $\rm{\dot{M}\,'=10^{7}\dot{M}}$ (Column 7) where $\rm{\dot{M}}$ is in solar 
 mass per year, the dust-to-gas abundance ratios $\rm{\rho _{d/g}=10^{4}\rho _{d}/\rho _{g}}$ (Column 8)
 and the expansion velocities $\rm{v_{e}}$ in $\rm{km\,s^{-1}}$ (Column 9).} 
\begin{flushleft}
{\scriptsize{
\begin{tabular}{ccccccccc}
\hline\noalign{\smallskip}
G & $\rm{T_{eff}}$ & l & m & C/O & P & 
$\rm{\dot{M}\,'}$ & $\rm{\rho _{d/g}}$ & $\rm{v_{e}}$ \\
\noalign{\smallskip}\hline
\noalign{\smallskip}
HC5 & 3470 & 1735 & 0.57      & 1.01      & 290     & 1.7 & 8.6 & 6.5 \\
    &      &      & $\pm$0.20 & $\pm$0.01 & $\pm$70 &     &     &     \\ 
CV1 & 3290 & 2265 & 0.55      & 1.005     & 299     & 5.4 & 9.2 & 6.7 \\
    &      &      & $\pm$0.15 & $\pm$0.01 & $\pm$98 &     &     &     \\ 
CV2 & 3095 & 4130 & 1.0       & 1.01      & 339     & 5.2 & 6.3 & 7.1 \\
    &      &      & $\pm$0.2  & $\pm$0.01 & $\pm$71 &     &     &     \\ 
CV3 & 2920 & 4885 & 1.6       & 1.03      & 355     & 2.7 & 5.0 & 7.3 \\
    &      &      & $\pm$0.3  & $\pm$0.03 & $\pm$73 &     &     &     \\ 
CV4 & 2775 & 5960 & 1.9       & 1.13      & 353     & 3.7 & 5.7 & 8.8 \\
    &      &      & $\pm$0.35 & $\pm$0.13 & $\pm$55 &     &     &     \\ 
CV5 & 2640 & 6530 & 2.6       & 1.2       & 389     & 2.7 & 5.8 & 9.8 \\
    &      &      & $\pm$0.5  & $\pm$0.2  & $\pm$40 &     &     &     \\ 
CV5 & 2645 & 1600 & 0.5:      &           & 423:    &     &     &     \\
und &      &      &           &           &         &     &     &     \\ 
CV6 & 2435 & 8635 & 4.2       & 1.4       & 448     & 2.5 & 6.9 & 13.4 \\
    &      &      & $\pm$0.8  & $\pm$0.4  & $\pm$53 &     &     &     \\ 
CV7 & 1950 & 16445 & 8.2:     & 1.4:      & 444:    & 8.7 & 5.3 & 13.9 \\
    &      &       &          & $\pm$0.4  & $\pm$107 &     &     &     \\ 
CV7 & 1950 & 16445 & 4.2      & 1.4:      & 444:    & 62  & 8.4 & 14.5 \\
\noalign{\smallskip}\hline
\end{tabular}
}}
\label{mass}
\end{flushleft}
\end{table}
   The circumstellar envelope characteristics of 9 carbon stars were probed by Scho\"{\i}er et al. 
   (\cite{schoie02}), combining radio measurements and infrared observations from ISO. Their radiative 
   transfer analysis allowed us to put constraints on the mass loss history of those objects, specially 
   CW Leo=IRC+10216 and RW LMi=CIT6 (studied here) and IRAS 15194-5115=II Lup=CGCS 3592 (not included). 
   They found that those stars are not likely to have experienced any drastic long-term mass loss rate
   modulations, at least less than a factor of about 5, over the past thousands of years. This is a 
   situation in contrast with the case of the detached shells mentioned above. 
 
   If we remove the nine symbols of those high mass loss
   objects in Fig.$~\ref{pmvste},$ the remaining filled circles (semi-regulars) and crosses (Miras)
   populate a locus with a flat portion in the 2900-2400 K range and for n=66 stars,
   \begin{equation} 
    \rm{ \log \dot{M} \simeq \left(-0.8\pm 1.7\right)\log T_{eff}-\left(3.6\pm 5.8\right)}
   \end{equation}
   with an average value of
   \begin{equation} 
    \rm{ < \dot{M} >\simeq \left(5.8\pm 4.4\right)\,10^{-7}\;\;M_{\sun}yr^{-1}}
   \end{equation}
   for this wide $\rm{ 10^{-7}-2\,10^{-6}\,M_{\sun}yr^{-1}}$ nearly horizontal strip. It is flanked by two 
   ranges of increasing mass loss  rate with decreasing effective temperature. For $\rm{T_{eff} \ge 2900\,K}$
   we obtain for 32 stars
   \begin{equation}
    \rm{ \dot{M} \simeq 4.51\,10^{20}\;T_{eff}^{-7.83}\;\;M_{\sun}yr^{-1}}
   \end{equation}
   and for $\rm{T_{eff} \le 2400\,K}$ with 17 stars
   \begin{equation}
    \rm{ \dot{M} \simeq 8.46\,10^{24}\;T_{eff}^{-9.20}\;\;M_{\sun}yr^{-1}}\;\;\;\;\;.
   \end{equation}
   Those three relations which are not satisfied by the 9 objects with strong mass loss and possibly a few 
   others, are not shown in Fig.$~\ref{pmvste}.$ Instead the predictions of Sect. 4 as quoted 
   in Table 4 are plotted as filled square-symbols connected by a full broken line. A very good agreement 
   is observed in the 3000-2400 K range of effective temperatures and the observed flat portion is
   confirmed by the predictions. We shall identify its cause later in this section. Below 2400 K,
   the observed locus is well reproduced by predictions provided $\rm{M=4.2\,M_{\sun}}$ is adopted
   instead of the uncertain $\rm{M=8.2\,M_{\sun}}$ value estimated by Bergeat et al. (\cite{berge02c})
   from a few extreme objects. This high value was possibly induced by overestimating the ``photospheric''
   radius of very cool giants with dust shells. The right portion of Fig.$~\ref{pmvste}$ is an additional
   argument in favor of masses not exceeding about $\rm{4\,M_{\sun}}$ for a large majority of luminous
   carbon-rich giants. For effective temperatures higher than 3000 K, there is an increasing disagreement
   between observations and predictions up to 3400 K. Hot variables exhibit an irregular pulsational
   behaviour with small amplitudes, resulting in dubious pulsation mode and period. Uncertainties increase
   accordingly. It should be noted that, according to Arndt et al. 
   (\cite{arndt97}), one has to refrain from applying Eqs. (4) to (6) beyond 3000 K since none of their
   48 models lies there. In addition, none of their models is fainter than $\rm{5\,10^{3}\,L_{\sun}}$
   contrary to the observed averaged mean luminosities of HC5, CV1 and CV2-stars in Table 4.
   We however attempted to apply the $\rm{<M>=1.6\,M_{\sun}}$ mean value of the CV3-group in Table 4 to
   the other earlier groups CV1 and CV2 with other data otherwise unchanged. Together
   with an additional calculation from Eq. (4) assuming for CV1 the $\rm{<M>=1.0\,M_{\sun}}$ value quoted
   for CV2 in Table 3, they are plotted as filled triangles in Fig.$~\ref{pmvste}$. The dashed
   lines connecting the CV3-square to the filled triangles for CV1 and CV2 and the initial segmented full 
   line reasonably well encompass the observations. 
   This result suggests that the slope for $\rm{T_{eff} > 2900\,K}$ in Fig.$~\ref{pmvste}$ is 
   indicative of a nearly uniform value of about one solar mass. A large correction
   factor (3.9 or 2.5) from Eq. (2) was applied to part of the contributing data. Without that correction,
   the shift of observations below the continuous line of predictions would have been substantially larger.
   \begin{figure*}
   \centering
   \resizebox{14.0cm}{!}{\includegraphics{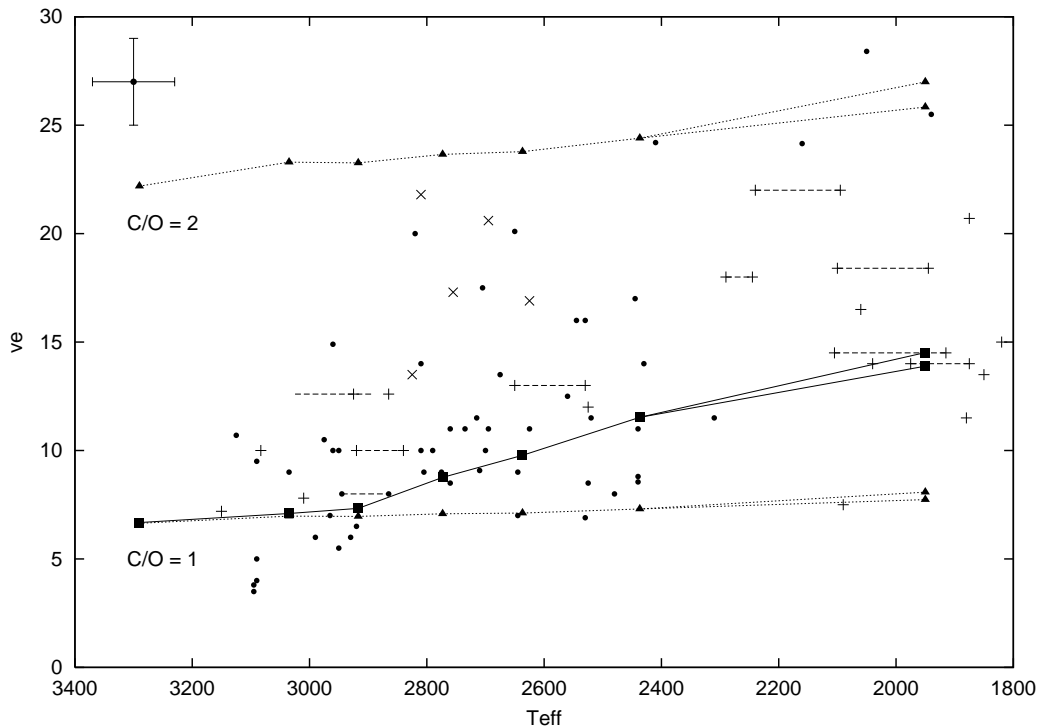}}
   \caption{Outflow (expansion) velocities in km/s as a function of effective temperatures in Kelvin, as
      taken from Table 2 (observations) and Table 4 (predictions from models). Typical error bars are also
      displayed at left. Same symbols used as in Fig.$~\ref{pmvste}.$ with 5 non-Miras with detached
      shells $\left(\times \right).$ The predictions from theory are plotted as filled squares connected by
      full lines. Extreme lines corresponding to C/O=1 and C/O=2 respectively are also shown (see text).}
    \label{vevsste}%
    \end{figure*}
   Similarly the slope below 2400 K seems
   to correspond to a nearly uniform value of about $\rm{4\,M_{\sun}}.$ The latter is predicted as
   the upper limit from evolutionary calculations presumably due to hot bottom-burning most effective in
   AGB stars of larger masses (Bergeat et al. \cite{berge02b} and references therein). We also calculated
   the mass loss rate of the CV6 and CV7-groups, assuming the $\rm{<M>=2.6\,M_{\sun}}$ value quoted for CV5
   in Table 3, keeping the other parameters unchanged. They are shown as filled triangles connected to
   the CV5-square by a broken dashed line. A possible explanation of higher mass loss rate might be
   smaller mass due to previous strong mass loss. Precisely, Scho\"{\i}er et al. (\cite{schoie02}) deduced
   that mass loss rates remained high over the past thousands of years in objects like CW Leo=IRC+10216 and
   RW LMi=CIT6. From the comparison of theoretical nucleosynthesis models and measurements of abundances in
   the circumstellar envelope, Kahane et al. (\cite{kahane00}) favor a mass not larger than 
   $\rm{2\,M_{\sun}}$
   for CW Leo, well below the $\rm{4.2\,M_{\sun}}-$limit considered for the stars of Eq. (10). Winters et al.
   (\cite{winte94a}) previously estimated $\rm{M\le 2\,M_{\sun}}$ with $\rm{C/O\simeq 1.4}$ for CW Leo. 
   {\it In other words, the apparent (lasting) ``superwinds'' could well be standard mass loss in objects
   of diminished masses.\/} 
   It should be noted that Arndt et al. (\cite{arndt97}) did not calculate models cooler than 2300 K.
   Additional data and models are clearly necessary before a firm conclusion can be reached.

   We shall see in Sect. 6 that the four stars
   with detached shells mentioned above also have expansion velocities larger than average. Actually,
   it happens that velocities are little influenced by adopted masses (Sects. 3 and 6). In addition,
   R Scl has now returned to low mass loss rates (Wong et al. \cite{wong04}). For those stars with
   detached shells and intermediate (2400-2900 K) effective temperature, the above explanation of enhanced 
   mass loss in terms of diminished masses is not acceptable.
   By continuity, we are lead to the conclusion that the 3000-2400 K flat portion corresponds to the 
   approximate $\rm{1-4\,M_{\sun}}$ mass range. Discussing their formula (1), i.e. our Eq. (4), Arndt et al. 
   (\cite{arndt97}) concluded that the abundance ratio, pulsation period and piston amplitude velocity
   have only small influences on the mass loss rate. Accordingly, we find no clear difference between the
   loci of Miras and semi-regulars. It remains to consider the effective temperature, the luminosity
   and the mass of the star. Since, on average, luminosity increases with decreasing effective temperature,
   both parameters contribute to an increase of the mass loss rate. Following Eq. (4), that increase is
   inhibited in the 3000-2400 K range by the increase of stellar mass in the $\rm{1-4\,M_{\sun}}$ domain.
   {\it The 3000-2400 K flat portion in Fig.$~\ref{pmvste}$ is a further argument in favor of the
   $\rm{1-4\,M_{\sun}}$ range for bright carbon-rich giants (less luminous objects can populate the
   $\rm{0.5-1\,M_{\sun}}$ domain), in addition to the use of evolutionary tracks in the HR diagram and to
   the calculations of present pulsation masses (Bergeat et al. \cite{berge02b}, \cite{berge02c}). We
   conclude that the models and relations from Arndt et al. (\cite{arndt97}) satisfactorily 
   reproduce the mass loss rate deduced from millimetric observations. Larger mean mass thus implies,
   other things being equal, a lower mass loss rate. \/}
\section{Outflow velocity {\it vs.} effective temperature}
   The outflow or expansion velocities $\rm{v_{e}},$ i.e. estimates of velocity at infinity in km/s as 
   quoted in Table 2 (Columns 8 and 17) are plotted in Fig.$~\ref{vevsste}$ as a function of the effective 
   temperatures from Columns 3 and 12. 
   The main trend of increasing expansion velocity with decreasing 
   effective temperature is observed as expected from theory (Sect. 3), but with a large scatter here. 
   Typical error bars on both coordinates are also shown and are much smaller.
   The same symbols as in Fig.$~\ref{pmvste}$ are used.
   The $\times$-symbols point to five stars with detached shells (U Ant, S Sct, R Scl, 
   U Cam and TT Cyg). The latter also show large velocities (14-21 km/s) for 
   $\rm{T_{eff} \simeq 2600-2800\,K}$ about a factor 2 larger than predicted values. We already found in 
   Sect. 5 that they display higher mass loss rates than predicted. The issue is dubious for the five 
   cool extreme objects of Sect. 5. They are scattered with CW Leo=IRC+10216 on the predicted relation
   and FX Ser located well above it.
   
   The predictions of $\rm{v_{\infty}}$ from Eq. (6) as quoted in Table 4, are shown in 
   Fig.$~\ref{vevsste}$ as filled square-symbols connected by a broken full line. The latter fits
   reasonably well the 86 \% of the stars in the sample that are located in the lower part of the 
   diagram. The models
   seem to correctly predict the low expansion velocities which roughly correspond to the low mass loss
   rates of Sect. 5, themselves well-predicted by the same models.
   The decreasing effective temperature and increasing luminosity along the sequence of Table 3 both
   contribute to the increase of the outflow velocity but the effect is small when compared to the 
   influence of the $\rm{\epsilon _{C}/\epsilon _{O}}$ abundance ratio. This point is illustrated by the two 
   dashed lines in Fig.$~\ref{vevsste}$ as obtained keeping the ratio to 1 or 2 instead of the mean values of
   Table 4. Within errors, the observed values lie between those two lines. This is precisely the range
   of the abundance ratio as derived from spectra and model atmospheres of the cool carbon-rich giants
   (Lambert et al. \cite{lamber86} and references cited in Sect. 2; see also Fig. 7 of Bergeat et al.
   \cite{berge02c}). 
   
   We also note that above 2900 K the observations in Fig.$~\ref{vevsste}$ tend to
   concentrate towards the C/O=1 line, below 10-15 km/s. This can be
   attributed to the fact that stars in the hotter domain of Table 4 do have mean abundance ratios close 
   to unity with a small range apart. Below 2900 K (CV3 and later groups), the C/O mean values and
   ranges increase (Table 4). On average, larger velocities are reached there. The influence of the
   C/O ratio is responsible for a triangular void in Fig.$~\ref{vevsste}$ at $\rm{T_{eff}\ge 2800\, K}$ 
   that is inserted between the C/O=1 and C/O=2 lines. {\it Predictions are consistent with millimetric
   observations and visible and infrared spectroscopy with $\rm{1\le \epsilon _{C}/\epsilon _{O} \le 2},$ 
   and with the mean values and dispersions as functions of effective temperature from Bergeat et al.
   (\cite{bergea01}), as displayed in Fig. 7 of Bergeat et al. (\cite{berge02c}). \/}
\section{Dust-to-gas ratio {\it vs.} effective temperature}
   The diagram of the dust-to-gas ratio against effective temperature is not shown here. There is
   actually no marked slope when Eq. (5) and the corresponding predictions of Table 4 are used. 
   In the 
   2300-3000 K validity domain of models from Arndt et al. (\cite{arndt97}), the mean value amounts to
   \begin{equation}
    \rm{<\rho _{d/g}>=<\rho _{d}/\rho _{g}>\simeq \left(5.9\pm 0.7\right)\,10^{-4}}
   \end{equation}
   As described in Sect. 2, the mean observed values quoted in Table 2 are typically ``lower'' ones
   derived assuming low drift velocities. For 72 documented stars, we obtained
   \begin{equation}
    \rm{\dot{M}_{d}/\dot{M}_{g}\simeq \left(1.3\pm 1.1\right)\,10^{-3}}
   \end{equation}
   a value about 2.2 times larger than the predicted one. There is practically no slope in the
   diagram of ``observed'' dust-to-gas ratios of rates against effective temperatures from Table 2,
   at least within the large error bars. It must be emphasized here that the data used was corrected
   for the f-factor of Sect. 2 and for a smaller factor for adopted luminosity
   \begin{equation}
    \rm{\left(\dot{M}_{d}/\dot{M}_{g}\right)_{BKR}=\left(\dot{M}_{d}/\dot{M}_{g}\right)_{Lit}\:
   \left(L_{BKR}/L_{Lit}\right)^{-0.5}}
   \end{equation}
   Without that f-correction, the above discrepancy ratio would have jumped from 2.2 to about 6.
   The mean ``upper'' value of $\left(3.2\pm2.5\right)\,10^{-3}$ derived in Sect. 2 is about
   5.4 times larger than the Eq. (11) predicted value. Apparently, adopting larger drift velocities could 
   make the discrepancy between observations and predictions even worse, but predicted values would also
   be increased. Very recent models with the influence of the grain drift velocity properly treated
   (Sandin \& H\"{o}fner \cite{sandin04}) can contribute to shift predictions closer to observations.
   The derivation of the ratio from the observations makes use of data from both infrared and millimetric
   ranges, with simpliflying assumptions (like a unique temperature for instance). It is not clear whether
   the same layers are involved in the extended atmosphere and shell (from inner dust condensation zone
   to outer regions with terminal velocity reached).

   Following Eq. (5) taken from Arndt et al. (\cite{arndt97}), it appears that in addition to 
   decreasing effective temperature and increasing luminosity whose influences counteract here, the 
   stellar mass and the $\rm{\epsilon _{C}/\epsilon _{O}}$ abundance ratio do have an appreciable 
   influence. The predicted dust-to-gas density ratio increases with both decreasing stellar mass and 
   increasing abundance ratio. Adopting for instance $\rm{M=1\,M_{\sun}}$ and 
   $\rm{\epsilon _{C}/\epsilon _{O}\ge 2}$ can rise the Eq. (11) value higher than that in Eq. (12), 
   reaching the level of the highest ``lower'' values observed. It seems outside the range of acceptable
   parameter values taking into account the analyses of Sects. 5 and 6 from Table 4 data.
   However, keeping the mass range of Table 4 that fits nicely the gas mass loss rate data (Sect. 5),
   an increase of about 20\% of the C/O ratios from Lambert et al. (\cite{lamber86}) can help, which
   remains compatible with the velocity comparison in Fig.$~\ref{vevsste}.$ Their abundance ratios for
   30 carbon stars lead to $<\rm{\left(^{12}C+^{13}C\right)/H}>\simeq\left(5.5\pm1.6\right)\,10^{-4}$
   in numbers of atoms and $\left(2.7\pm1.0\right)\,10^{-3}$ in mass for a gas principally of H and He, 
   if their C/O ratios are kept.
   The estimate is then of 22\% of carbon locked up in grains if prediction of Eq. (11) is trusted
   against 48\% for the mean observed ratio of Eq. (12). If confirmed, such figures would mean that 
   carbon molecules could be appreciably depressed. Very high drift velocities could lead to 
   unreasonably high estimates.
\section{Discussion and conclusions}
   We have compared the results of millimetric (and infrared) observations, namely gas mass loss rate, 
   expansion (outflow) velocity and dust-to-gas density ratio, to the predictions from hydrodynamical
   time-dependent models of dust-driven winds for carbon-rich chemistry. Corrections were applied to
   the published ``observed'' values of the first (distance and radiative transfer) and the third 
   quantity (luminosity and radiative transfer) as described in Sect. 2.
   The prediction of models are applied to long
   period variables whose pulsation is simulated by a piston of given amplitude velocity and period 
   acting at the basis of the atmosphere. We made use of the approximate formulae of Arndt et al.
   (\cite{arndt97}) based on 48 models exploring ranges in the six parameters they include. Among 
   them, the pulsation period and the piston amplitude velocity have only little influence (Sect. 3).
   We found no marked difference in the observed data between Miras and non-Miras
   (here semi-regulars with available periods). Four parameters remain whose ranges may
   influence the results, namely effective temperature taken as the stellar temperature used in the 
   models, luminosity, mass and carbon to oxygen abundance ratio. The observational
   data was summarized in Table 2 for 119 carbon-rich giants formerly studied by Bergeat et al.
   (\cite{bergea01}, \cite{berge02b} and \cite{berge02c}), for effective temperatures, luminosities,
   stellar masses and C/O abundance ratio. The formulae (4), (5) and (6) as taken from Arndt et al.
   lead for the mean values of Bergeat et al. to the estimated means of Table 4 for the samples 
   corresponding to the photometric groups HC5 and CV1 to CV7 with a range of effective temperatures
   from 3470 K down to 1950 K. This latter parameter is considered as the most accurate one at 
   present and it is used for reference. We also hold fixed to reference values the stellar mass
   and the C/O abundance ratio keeping unchanged the run of effective temperatures and luminosities
   of Table 4, namely the TP-AGB of Galactic carbon-rich giants in the HR diagram, as obtained by 
   Bergeat et al. (\cite{berge02c}).

   As described in Sect. 5, we obtained a good agreement between predictions and observations in the
   diagram of mass loss rate {\it vs.} effective temperature (Fig.$~\ref{pmvste}$). For about 90\% of
   our sample, the mechanism involving gas lifted by pulsation and radiation pressure on condensed 
   dust is confirmed. Three equations (7) and thus (8), (9) and (10) were obtained in the three ranges 
   of effective temperatures distinguished. The usual statement (van Loon et al. \cite{vanloo03}) that
   $\rm{\dot{M}\propto T_{eff}^{-8}}$ is approximately true only above 2900 K (Eq. (9)) and below 
   2400 K (Eq. (10)), with different proportionality coefficients however. There is practically no 
   variation over the 2400-2900 K range in Fig.$~\ref{pmvste}$ (including about 55\% of our sample).
   {\it Predictions and observations are in agreement provided the variations 
   of luminosity on the AGB in the HR diagram (Bergeat et al. \cite{berge02b}) and the mean 
   masses ranging from about $\rm{1M_{\sun}}$ around 3000 K to about $\rm{4M_{\sun}}$ around 2400 K
   (see Table 4), as deduced from theoretical tracks in the HR diagram and from pulsations masses
   (Bergeat et al. \cite{berge02c}).\/} Here mass is the third important parameter, with influence much
   larger than that of the C/O abundance ratio. At low effective temperatures, Eq. (10) corresponds
   to a constant mass of about $\rm{4M_{\sun}}.$ At high temperatures, Eq. (9) describes carbon
   variables of about $\rm{1M_{\sun}}$ or less, but both observations and predictions are uncertain 
   there. Nine giants in our sample (at least) do have mass loss rates larger than predicted from Eqs.
   (7) to (10). Four cases of detached shells are found in the 2600-2800 K range, i.e. objects having 
   experienced highly episodic mass loss possibly caused by a thermal pulse. One of them (R Scl) has
   returned to a moderate mass loss rate (Wong et al. \cite{wong04}), and this is certainly
   true for the others (e.g. Olofsson et al. \cite{olofss00} for TT Cyg).
   This is a strong argument in favor of a specific mass loss mechanism at work. The other five stars
   are very cool objects with $\rm{T_{eff}\le 2500 K}$ and optically-thick dust shells shown by
   strong infrared excesses, a consequence of very high mass loss rates, like RW LMi (CIT6) or CW Leo
   (IRC+10216). Scho\"{\i}er et al. (\cite{schoie02}) have shown that the gas mass loss rate of those
   extreme objects did not vary much over the last thousand of years. As argued in Sect. 5, they
   may be stars with masses of about $\rm{1.5-2.5M_{\sun}},$ in contrast with the $\rm{4.2M_{\sun}}$ 
   favored for the cool objects with lower rates as given by Eq. (10). If true, no specific mass
   loss mechanism would be required here, contrary to the case of detached shells.    
     
   A good agreement is also obtained between predictions and observations in the diagram of expansion
   (outflow) velocity {\it vs.} effective temperature (Fig.$~\ref{vevsste}$), as described in Sect. 6. 
   The expansion (outflow) velocity increases with decreasing effective temperature, with about 
   86 \% of the objects within the range of predictions as quoted in Table 4 from Eq. (6). 
   To within error bars, the observations are located between the two lines obtained for 
   $\rm{\epsilon _{C}/\epsilon _{O}=1}$ and $\rm{\epsilon _{C}/\epsilon _{O}=2}$ respectively, the
   other parameters being kept equal to their values in Table 4 (including the mean mass distribution
   successfully used for mass loss rates and described above). This is the range deduced from
   spectroscopy and model atmospheres (Lambert et al. \cite{lamber86}; see also Fig. 7 of Bergeat et 
   al. \cite{berge02c}). In Fig.$~\ref{vevsste},$ it is remarkable that for effective temperatures 
   higher than about 2900 K, the stars concentrate towards the $\rm{\epsilon _{C}/\epsilon _{O}=1}$ 
   line, below 10-15 km/s. The maximum observed velocity rapidly decreases with increasing 
   temperature, leaving empty a large triangular area in Fig.$~\ref{vevsste}.$ The result of Bergeat
   et al. (\cite{bergea01}, Fig. 7 in \cite{berge02b}) is confirmed that above 2900 K, the 
   C/O ratio remains close to unity with little dispersion. 
   The $\rm{\epsilon _{C}/\epsilon _{O}}$ abundance ratio is confirmed here as the third major 
   parameter with effective temperature and luminosity, while stellar mass has little influence in
   Fig.$~\ref{vevsste}.$ Cases of superwind however blur the information from that figure.
   {\it We conclude that both mass loss rates and outflow velocities are correctly modeled from
   recent hydrodynamical calculations of dust-driven winds. Both the levels and the dependence 
   on effective temperature are fairly well reproduced. \/}
 
   Both the predicted and observed ratios of dust-to-gas mass loss rates show little variation with
   effective temperature if any. As found in Sect. 7, the average observed value of
   $\rm{<\dot{M}_{d}/\dot{M}_{g}>\simeq \left(1.3\pm 1.1\right)\,10^{-3}}$ is a factor 2.2 larger
   than the mean predicted value 
   $\rm{<\rho _{d/g}>=<\rho _{d}/\rho _{g}>\simeq \left(5.9\pm 0.7\right)\,10^{-4}}$ from Eq. (5).
   We have discussed in Sect. 7 the influence of the stellar mass and C/O abundance ratio. It
   should be noted that the uncertainties are very large here. Among the possible explanations is a
   moderate underestimate of abundance ratios (say 20\%) from model atmospheres, the overal agreement 
   of Fig.$~\ref{vevsste}$ being preserved. Another explanation of the above-mentioned disagreement
   might be some phenomenon in the micro-physics of dust grains, especially the non-equilibrium dust
   formation process (Andersen et al. \cite{anders03}, Sandin \& H\"{o}fner \cite{sandin04} and 
   references therein). Allowing drift in the time-dependent (grain growth) wind models alters their 
   structure. In some cases there is several times more dust in drift models than in non-drift ones,
   even at low drift velocity (a few km/s). It is also found that drift plays an active role in 
   accumulating dust to certain narrow regions (Sandin \& H\"{o}fner \cite{sandin04}), which renders
   the above comparison even more difficult. It was however seen in Sect. 7 that element abundances  
   from model atmospheres raise constraints limiting the magnitude of such effects. Further 
   investigations are needed before we can determine which estimate is closer to truth. 

   We finally give in Appendix A a few simple approximate formulae in a parametric form. They 
   reproduce the results of the present study with fairly good accuracy. They can be useful for
   readers wishing not to enter into the full details of the present analysis.
 \begin{acknowledgements}
   We gratefully acknowledge constructive criticism and valuable suggestions from an anonymous
   referee. Our analysis was improved and a few points clarified.
 \end{acknowledgements}

\appendix
\section{Approximate formulas}
   Some readers may wish not to enter into the full details of the present analysis. From Eq. 6 and
   Fig.$~\ref{vevsste},$ we propose for the expansion velocities
   \begin{eqnarray}
    \rm{\log v _{\infty} =\left(-0.219\pm0.012\right)\log T_{eff}+\left(1.603\pm0.041\right)}
    \nonumber\\
    \rm{+1.74\log \left(\epsilon _{C}/\epsilon _{O}\right)} \:\:\:\:\:\:\:\:
   \end{eqnarray}
   for $\rm{T_{eff}\le 3150\,K},$ and for $\rm{T_{eff}\ge 3150\,K},$
   \begin{eqnarray}
    \rm{\log v _{\infty} =\left(-0.682\pm0.056\right)\log T_{eff}+\left(3.219\pm0.198\right)}
    \nonumber\\
    \rm{+1.74\log \left(\epsilon _{C}/\epsilon _{O}\right)} \:\:\:\:\:\:\:\:
   \end{eqnarray}
   where the first two terms include the data in Table 4 concerning the other (minor) parameters.
   Concerning the mass loss rates (Eq. (4) and Fig.$~\ref{pmvste}$), we can write
   \begin{eqnarray}
    \rm{\log \dot{M} =\left(-14.62\pm0.18\right)\log T_{eff}+\left(44.66\pm0.60\right)}
    \nonumber\\
    \rm{-2.86\log \left(M/M_{\sun}\right)} \:\:\:\:\:\:\:\:
   \end{eqnarray}
   for $\rm{T_{eff}\le 2900,K},$ and for $\rm{T_{eff}\ge 2900\,K},$
   \begin{eqnarray}
    \rm{\log \dot{M} =\left(-19.12\pm0.44\right)\log T_{eff}+\left(60.21\pm1.53\right)}
    \nonumber\\
    \rm{-2.86\log \left(M/M_{\sun}\right)} \:\:\:\:\:\:\:\:.
   \end{eqnarray}
   For the dust-to-gas density ratio, the observations point to
   $\rm{<\rho _{d/g}>=<\rho _{d}/\rho _{g}>\simeq \left(1.3\pm 1.1\right)\,10^{-3}}$ 
   while predictions from Eq. (5) can be written as
   \begin{eqnarray}
    \rm{\log \left(\rho _{d}/\rho _{g}\right) =\left(1.630\pm0.268\right)\log T_{eff}+
    \left(-9.007\pm0.906\right)}\nonumber\\
    \rm{+2.45\log \left(\epsilon _{C}/\epsilon _{O}\right)} \:\:\:\:\:\:\:\:
   \end{eqnarray}
   for $\rm{T_{eff}\le 2775,K},$ and for $\rm{T_{eff}\ge 2775\,K},$
   \begin{eqnarray}
    \rm{\log \left(\rho _{d}/\rho _{g}\right) =\left(4.635\pm0.605\right)\log T_{eff}+
    \left(-19.35\pm2.10\right)}\nonumber\\
    \rm{+2.45\log \left(\epsilon _{C}/\epsilon _{O}\right)} \:\:\:\:\:\:\:\:.
   \end{eqnarray}
   The mean effective temperatures and luminosities (in solar units) of Table 4 follow the 
   relation
   \begin{equation}
    \rm{\log l =\left(-3.065\pm0.100\right)\log T_{eff}+\left(14.31\pm0.34\right)}
   \end{equation}
   for $\rm{T_{eff}\le 3050,K},$ and for $\rm{T_{eff}\ge 3050\,K},$
   \begin{equation}
    \rm{\log l =\left(-6.528\pm0.693\right)\log T_{eff}+\left(26.34\pm2.43\right)}
   \end{equation}
   two relations where the true dispersion in the H-R diagram is not taken into account. The
   mass-luminosity relation in Table 4 was described in detail in Sect. 6 of Bergeat et al.
   (\cite{berge02c}; their equations (10) to (12)). In solar units, we obtain
   \begin{equation}
    \rm{\log l =\left(0.72\pm0.07\right)\log m\,+\left(3.58\pm0.12\right)}
   \end{equation}
   for the whole range $0.5-4.2\,M_{\sun}$ or 3500-2300K. Restricting ourselves to the 
   best-documented
   region from CV2 to CV5 (3150-2550K) roughly corresponding to $1-3\,M_{\sun},$ we derive
   \begin{equation}
    \rm{\log l =\left(0.497\pm0.075\right)\log m\,+\left(3.658\pm0.034\right)}
   \end{equation}
   that is nearly
   \begin{equation}
    \rm{l \propto\ m^{1/2}}
   \end{equation}

\end{document}